\documentclass[amsmath,amssymb,preprint,prl]{revtex4}

\begin{document}

\setlength{\unitlength}{1mm}
\textwidth 15.0 true cm
\textheight 22.0 true cm
\headheight 0 cm
\headsep 0 cm
\topmargin 0.4 true in
\oddsidemargin 0.25 true in
\thispagestyle{empty}

\def\beq{\begin{eqnarray}}
\def\eeq{\end{eqnarray}}

\def\pr{\partial}
\def\Mpl{M_{\rm Pl}}
\def\Rbr{R}
\def\Rbu{{\cal R}}
\def\M{M_*}
\def\gn{G_N}
\def\tgn{{\tilde G}_N}
\def\goesto{\rightarrow}
\def\eps{\epsilon}
\def\x{\bf x}

\title{Photon Mass Bound Destroyed by Vortices}

\author{Eric Adelberger$^{a}$, Gia Dvali$^b$, Andrei Gruzinov$^{b}$}

\address{$^a$ Center for Experimental Nuclear Physics and Astrophysics, University of Washington, Seattle, WA 98195-4290}

\address{$^b$ Center for Cosmology and Particle Physics, Department of Physics, New York University, NY 10003}

\date{June 13, 2003}

\setcounter{footnote}{0} \setcounter{page}{1}
\setcounter{section}{0} \setcounter{subsection}{0}
\setcounter{subsubsection}{0}

\begin{abstract}

The Particle Data Group gives an upper bound on the photon mass  $m  < 2 \times 10^{-16}$eV from a laboratory experiment and lists, but does not adopt, an astronomical bound $m < 3 \times 10^{-27}$eV, both of which are based on the plausible
assumption of large galactic vector-potential. We argue that the interpretations of these experiments should be changed, which 
alters
significantly the bounds on $m$.
If $m$ arises from a Higgs effect, both  %
limits are
invalid
because 
the Proca vector-potential of the
galactic magnetic field may be neutralized by 
vortices giving a large-scale magnetic field that is effectively Maxwellian.
In 
this regime, experiments 
sensitive to
the Proca potential do not yield a useful bound on $m$. As a by-product, the non-zero photon mass from Higgs effect predicts generation 
of a primordial magnetic field in the early universe. 
If, on the other hand, the galactic magnetic field is in the Proca regime, the very existence of the observed large-scale magnetic field gives $m^{-1}\gtrsim 1$kpc, or $m\lesssim 10^{-26}$eV.  
 
\end{abstract}
\maketitle

%
{\bf INTRODUCTION:} The possibility of a non-zero photon mass remains one of the
most important issues in physics, as it would shed
light on 
fundamental questions such as 
charge conservation, charge quantization, the possibility
of the charged black holes and magnetic monopoles, etc.
%
The most stringent upper bounds on the photon mass listed by the
Particle Data Group \cite{PDG}, $m<3\times 10^{-27}$ eV 
and  $m<2\times 10^{-16}$,
are based on the
assumption that 
a massive photon would cause
large-scale magnetic fields to be accompanied by an energy density
\begin{equation}
 m^2_A \, \tilde{A}_{\mu}\tilde{A}^{\mu}
\label{Procaenergy}
\end{equation}
associated with the Proca field $\tilde{A}_{\mu}$ that describes the massive photon\cite{gold}.
This manifests itself in two different ways.
The first limit comes from the potential
astrophysical effects \cite{yam, chibisov}, and
the second 
from an experiment that
used a toroidally magnetized pendulum \cite{lakes} to measure the magnetic
field gradient in a magnetically shielded vacuum. A recent experiment\cite{luo}
using an improved technique
obtained a similar result.
Both experiments actually measured the product $m^2\tilde{A}$,
where the ambient Proca vector potential is presumably dominated by the
field
of the galaxy. The value 
assumed in \cite{lakes,luo}, $\tilde{A}\sim R B
\sim 1\mu $G$\times$kpc, is astronomically reasonable as the large-scale,
$R\sim 1$kpc, galactic field has a strength $B\sim 1\mu$G.

Let us review
the standard arguments behind these bounds,
which assume that a massive photon
at low energies is described by the Proca field \cite{gold}. (Throughout this paper we denote
the Proca field as $\tilde A_{\mu}$, whereas $A_{\mu}$ should be understood as the Maxwellian
field.)  The Lagrangian density for $\tilde{A}_{\mu}$ is
\begin{equation}
{\cal L} \, = \,  -{1 \over 4} F_{\mu\nu}F^{\mu\nu} \, + \, {1 \over 2} m^2_A \, \tilde{A}_{\mu}\tilde{A}^{\mu} \, + 
\, \tilde{A}_{\mu}J^{\mu}
\label{Proca}
\end{equation}
where $F_{\mu\nu}$ is the usual field strength, and $J^{\mu}$ is the conserved current. 
The $m_A^2$-term in Eq. \ref{Proca}, makes $\tilde{A}_{\mu}$
experimentally observable. 
Naively, the existence of the Galactic magnetic field ${\bf B}_{\rm gal}\, = \,
{\nabla} \times {\bf \tilde{A}}_{\rm gal}$, implies an ambient
galactic vector potential 
\begin{equation}
\tilde{A}_{\rm gal} \, \sim \, B_{\rm gal}\, R_{\rm gal},
\end{equation}
The associated Proca energy $m^2_A \, {\bf \tilde{A}}^2$ can be detected by direct 
or indirect observations.
Indirect observations
rely on the effect the Proca energy would have on the galactic plasma \cite{chibisov},
implying the limit  $m_A < 3 \times 10^{-27}$eV.
The direct detections \cite{lakes} are based on measuring the torque on a magnetized ring, 
which depends on the angle between $\tilde{A}_{\rm gal}$ and the vector potential 
of the 
ring,
because the energy density contains a term 
$\sim m_A^2{\bf \tilde{A}}_{\rm gal}\cdot {\bf \tilde{A}}_{\rm ring}$~.
The null result \cite{lakes} implies 
the limit $m_A \leq 2 \times 10^{-16}$eV. 

We claim that photon 
mass bounds cannot be  established {\em without specifying
the microscopic origin of the mass}. In particular, if $m$
arises from the commonly accepted Higgs mechanism, the above bounds do not apply over a large portion of the parameter space.
It is quite possible for large-scale magnetic fields to be effectively Maxwellian, even if photons are massive. In this case observations of large-scale fields, say from the galaxy or from Jupiter, are not sensitive to $m$. This leaves us with the upper bound from laboratory tests of Coulomb's law, $m\lesssim 10^{-14}$eV \cite{will}.
%
It is also possible that the large-scale fields do remain in the Proca regime. But then, the available information about the large-scale magnetic field of the galaxy, and the gas pressure in the galaxy, actually gives a much more stringent bound, $m^{-1}\gtrsim R\sim 1$kpc, or $m\lesssim 10^{-26}$eV \cite{yam, chibisov}. 
%

{\bf THE HIGGS SCENARIO: }If photon has a non-zero mass, there are excellent field-theoretic
arguments for thinking it should arise from a
Higgs-type effect, 
in which case the above mass limits are invalid.
Note that when estimating the Proca field associated with the
galactic magnetic field, one 
must distinguish the actual
Proca field $\tilde{A}_{\mu}$, which can be measured, from its Maxwellian component
$A_{\mu}$;
$\tilde{A}_{\mu}$ has 3 physical degrees of freedom
(polarizations) as opposed to $A_{\mu}$ , which has only two. So 
that by giving a mass term to  $A_{\mu}$, we are supplementing it with an additional degree 
of freedom. The Proca field can be written as
\begin{equation}
\label{rad}
\tilde{A}_{\mu} \,= \, A_{\mu} \, - \, {1 \over g} \partial_{\mu} \psi~,
\end{equation}
where $\psi$ is the additional (longitudinal) polarization. Written in this way,
the Proca theory is manifestly gauge invariant under:
\begin{equation}
\label{rad1}
 A_{\mu} \, \rightarrow \,  A_{\mu}\, + \, {1\over g}\partial_{\mu}\,
\omega,
~~~
\psi \rightarrow \, \psi\, + \,\omega
\end{equation}
The 
new degree of freedom enters the Proca action only in the mass term; it cancels 
in $F_{\mu\nu}$ as well as in the couplings to the conserved current. 
When computing the galactic Proca field, we must be sure it is not 
compensated by the additional polarization. This is, in fact, what happens in the Higgs 
scenario.

Proca theory, Eq. \ref{Proca}, can be extended to the Higgs theory, by
promoting $m_A$ into a real scalar field $\phi = m_A/g$, with the self-Lagrangian
\begin{equation}
\label{phi}
{1 \over 2}(\partial_{\mu}\phi)^2 \, - \, {\lambda^2 \over 2} (\phi^2 \, - \, \eta ^2)^2.  
\end{equation}
Then,  $\phi$ can be thought of as the modulus
of a complex Higgs field,
$ H \, = \, \phi\, {\rm e}^{i\psi}$,
and $\psi$ is its phase (Goldstone boson), which becomes a longitudinal photon. 
The static energy in the absence of the electric field is
\begin{equation}
{\cal E} \, = \, \int d^3x \, \left ( \, {B^2 \over 2} \, + \, {g^2 \over 2} \phi ^2 \left (
{\bf A} \, - \, {1 \over g} \nabla \psi \right )^2 \, + {1 \over 2} (\nabla \phi )^2\, + \, 
{\lambda^2 \over 2} (\phi^2 \, - \, \eta ^2)^2\right ) .
\end{equation}
The two important parameters of the theory
are the mass of the photon $m_A = g\eta$, and the mass of the Higgs
particle $m_{\phi}\, = \, \lambda\, \eta$.
With frozen $\phi = \eta$, the theory simply reduces to Proca theory with the energy
\begin{equation}
{\cal E} \, = \, \int d^3x \, \left ( {B^2 \over 2} \, + \, {g^2 \over 2} \eta^2
{\bf \tilde {A}}^2 \right ),
\end{equation}
where now $\tilde{A}$ carries three degrees of freedom, only two of which contribute into $B$.

Naively, one may think that as long as $\lambda ^2 \eta ^4$ is bigger than the
energy density of the galactic magnetic 
field $B^2$ (with $B \, \sim \, \mu$G $\sim 10^{-25}$ GeV$^2$),
one can ignore the fluctuations in the Higgs field, and the theory
should effectively reduce to Proca. 
This is not correct, because the system has a choice of lowering a huge Proca energy,
stored over a large volume, by locally exciting the Higgs field. Even if the Higgs field
is much heavier than the value of the magnetic field in question  
for a big portion of parameter space,  the local Higgs energy is in fact less
costly than the alternative Proca gauge-field energy. Crudely speaking, 
if $m$
is due to a Higgs effect, then the Universe is effectively a type II
superconductor
where magnetic fields create
Abrikosov vortices.

This effect can be readily demonstrated by taking an example 
with a constant magnetic field $B_z = 2B$. The corresponding Maxwellian vector potential
(up to gauge transformation) is
$A_{\theta} = Br$, and in the absence of a third polarization
could naively contribute a divergent Proca energy. However, this energy is canceled
by non-trivial winding of longitudinal photon.
The Proca energy density is
\begin{equation}
 g^2 |\phi|^2 \left (Br - {1 \over gr} \partial_{\theta}\psi \right )^2~,
\end{equation}
which cancels if on average
 $\partial_{\theta}\psi \, = \, g\, B\, r^2$~.
This configuration of $\psi$ is impossible in the Proca theory, but can occur in the Higgs case
if there is an uniform density of zeros of Higgs field $\phi$, around which the phase winds
non-trivially, producing vortices. The integral around a closed circle of radius $r$
\begin{equation}
{1 \over 2\pi}\int \partial_{\theta}\psi = N(r)
\end{equation}
defines a winding number $N(r)$ which is equal to the number of vortices located inside the
circle.
Around each vortex $\psi$ changes by $2\pi$. The system cancels the Proca energy by creating uniform density of vortices 
$ n = gB/\pi$~.
The cancellation cannot be exact because $N(r)$ is a discrete number, so
 the residual Proca  energy density is
$\sim g B\eta^2$~.  
Equating 
this to the Higgs energy,
$g B\eta^2 \sim  \lambda^2 \eta^4$, 
 we get a critical value of the magnetic field $B_c$
\begin{equation}
B_c \sim \lambda^2 \eta^2/g~. 
\label{bcritical}
\end{equation}
For $B > B_c$ 
it is energetically favorable for $\phi$ to vanish everywhere, and the theory becomes Maxwellian.
 The same value of $B_c$ can be obtained by requiring that the Higgs cores overlap. 
That is,
the inter-vortex distance becomes equal to the inverse Higgs mass:
\begin{equation}
{1\over \sqrt{n_c}} \sim  {1\over \sqrt{B_cg}} \sim {1 \over \lambda \eta} 
\label{critdistance}
\end{equation}


{\bf CRITICAL VALUES OF B: }There are several interesting critical values of the system parameters. 
The first is the 
magnetic field given by Eq. \ref{bcritical}. 
For   $B > B_c$  the Higgs VEV vanishes and the photon becomes massless. 
Even if the galactic magnetic field is above $B_c$ this limit can still be of interest,
because the extra-galactic magnetic field 
can be below $B_c$.
Then the photon will be massive outside the galaxy, but massless inside. In such a regime, the  information about $m$ can only come from extra-galactic observations. 

For $B<B_c$, the photon is massive everywhere, and there are two regimes: Proca and non-Proca.

The 
system can only be in the Proca regime
(i.e. vortex formation is unfavorable) when its size satisfies
$R \, \lesssim \, 1/\sqrt{gB}$. 
For the galaxy, assuming $m_A \sim 10^{-14}$eV and $\lambda = 1$, this limit requires $\eta \sim 10^{22}$GeV!

If
$R \, \gtrsim \, 1/\sqrt{gB}$, 
vortices are energetically favored, but two sub-regimes are possible. 
The first occurs when 
$R \, \gtrsim \, \lambda \eta/gB$.
Then the system {\it classically} creates vortices 
out of vacuum 
and neutralizes the Proca energy. For the galaxy, assuming $m_A \sim 10^{-14}$eV and $\lambda = 1$, this requires $g\sim 10^{-16}$ and $\eta \sim 10^{2}$eV. Such 
a light, weakly-charged Higgs is compatible with all existing experimental data and naturalness bounds. 
At each point, the typical number of magnetically ovelapping vortices is ${\tilde N}\sim m_A^{-2}gB$. If ${\tilde N} \,\gg \,1$, the field is effectively Maxwellian. For the galaxy, assuming $m_A \sim 10^{-14}$eV and $g\sim 10^{-16}$, we get ${\tilde N} \,\sim \, 10^{5}$. (Situations with 
${\tilde N} \, \sim \, 1$
could provide experimental signatures that would be smoking guns for the Higgs scenario.)

The opposite case occurs when
$R \, \lesssim \, \lambda \eta/gB$.
Vortices are still energetically favorable, but the system cannot create them classically so that 
their existence will depend on pre-existing conditions such as phase transitions
in the early Universe. Due to 
its very small charge, the Higgs field decoupled 
from ordinary matter very early 
so that a phase transition
with vortex formation could have preceded formation of the magnetic field.
The evolution of such vortices
is not yet understood,
but is expected to be different from more conventional cosmic-string
networks \cite{vilenkin}.

Although 
we focused on Proca-Higgs cases, our analysis can be extended to alternative
gauge-invariant, ghost-free theories of the photon mass \cite{dgs} in
which $E$-field of a point charge for $r \ll m^{-1}$ is not screened
but rather modified to a higher inverse power law $\sim 1/r^3$.
In such cases the constraints may be even milder.

{\bf PRIMORDIAL MAGNETIC FIELD:} As an interesting by-product, the non-zero photon mass naturally predicts generation 
of a self-sustained primordial magnetic field in the early universe. Indeed if photon aquires mass
by the Higgs mechanism, than the thermal phase transition in $\phi$ would
inevitably form vortices by the Kibble mechanism \cite{Kibble}.

Due to small charge,
$\phi$ is never in thermal equilibrium with the Standard Model
species, but
because of the large self-coupling, it is in thermal equilibrium with itself.
Thus, due to the usual high-temperature symmetry restoration, the expectation value of 
$\phi$ had to vanish at early times. The only situation in which
$\phi$ would not vanish, would be if it never was in a thermal
equilibrium since inflation. This is unlikely -- even if $\phi$ had no direct coupling with inflaton,  it would still be
produced gravitationally with a Gibbons-Hawking temperature
($T_{GH} \sim 10^{14} GeV$ for
the standard inflation), unless $m_{\phi} \, > \, T_{GH}$.

The vorticies are produced when the temperature of the $\phi$ field drops to $T_{\phi}\sim m_{\phi}$. The standard big bang nucleosynthesis
requires that the temperature in $\phi$ quanta be smaller than the temperature in the standard model sector, $T_{\phi}\, <
\,
T_{SM}$, and the vortex network would form before galaxies if 

\begin{equation}
\label{primodialbound}
m_{\phi} \, > \, {T_{\phi} \over T_{SM}} 10^{-3}{\rm eV}.
\end{equation}

Thus, formation of a primordial magnetic field is a direct consequence
of the photon mass. The evolution of this magnetic field after formation, and its role for the galactic dynamo will be discussed elsewhere \cite{progress}.

%
{\bf THE PROCA REGIME: } If the galaxy is in the Proca regime
the averaged magnetic pressure is (see below)
\begin{equation} \label{pressure}
p_{\rm magnetic}={{\bf B}^2\over 24\pi}-{m^2{\bf {\tilde A}}^2\over 24\pi}.
\end{equation}
(The electric pressure is much smaller because the interstellar medium is a good conductor). 
In a stable system, this magnetic pressure must be counterbalanced by the plasma pressure and/or the plasma kinetic energy. The interstellar medium of our galaxy is in approximate equipartition
 assuming conventional electrodynamics \cite{shu}; the kinetic energy density of the plasma, the plasma pressure, and the standard magnetic energy density $B^2/(8\pi )$ are comparable to each other. Therefore, the ``massive'' part of the full magnetic pressure cannot exceed the standard part
$m^2{\bf {\tilde A}}^2\lesssim {\bf B}^2$,
which, together with the estimate ${\tilde A}\sim RB$, gives our bound
\begin{equation} \label{bound}
m\lesssim R^{-1} \lesssim 10^{-26}~{\rm eV} .
\end{equation}

In fact, this upper bound, derived from energy equipartition, was already discussed by Yamaguchi \cite{yam}, but then dismissed because the energy source of the magnetic field is unknown \cite{gold}. But, no matter what the source of the energy, the interstellar medium of the galaxy must provide the pressure support against the anomalous negative magnetic pressure, and the Yamaguchi estimate is, in fact, correct.

We can repeat the above analysis in terms of the Lorentz force. In Proca theory, one can still calculate magnetic fields from Ampere's law, if a new current density $m^2{\bf {\tilde A}}$ is added to the usual electric current ${\bf j}$. Then, approximately, $B\sim Rj+m^2R^2B$. Assume for the moment that 
$m$ is equal to the PDG upper bound, $\sim 10^{-16}$eV. Then $m^2R^2>>1$, and the usual current must precisely balance the ``massive current'', $4\pi {\bf j}=m^2{\bf {\tilde A}}$. But this current results in a huge Lorentz force density, $jB$, which cannot be possibly counterbalanced by the large-scale pressure gradients or accelerations. 

Finally, the same upper bound can be presented as a virial theorem \cite{chibisov}. The virial theorem relates mean values of different forms of energy for a system of particles and fields executing a bound motion. For the Proca theory, the energy-momentum tensor is
\begin{equation} 
T^{\mu }_{\nu }={1\over 4\pi }\left( -F^{\mu \alpha }F_{\nu \alpha }+{1\over 4}\delta ^{\mu }_{\nu }F^2\right) +{m^2\over 4\pi }\left( {\tilde A}^{\mu }{\tilde A}_{\nu }-{1\over 2}\delta ^{\mu }_{\nu }{\tilde A}^2\right).
\end{equation}
The magnetic pressure is defined as the magnetic part of $T_i^i$, and gives Eq. \ref{pressure}. The virial theorem, proved along standard lines \cite{ll}, is
\begin{equation} 
{\cal E} = -T-{m^2\over 4\pi }\int d^3r \langle {\tilde A}^2 \rangle,
\end{equation}
where $\cal{E}$ is the energy of the system, $T$ is the mean kinetic energy of the particles, $\langle {\tilde A}^2\rangle=\langle\Phi ^2\rangle-
\langle{\bf {\tilde A}}^2 \rangle$ is the mean squared 4-potential. Assuming that $B$-fields are much larger than the $E$-fields, we 
obtain
\begin{equation} \label{}
-U_g+{m^2\over 8\pi }\int d^3r \langle{\bf {\tilde A}}^2 \rangle=2T+{1\over 8\pi }\int d^3r \langle{\bf B}^2 \rangle,
\end{equation}
which shows that, for virialized motion, kinetic energy (which includes plasma pressure and kinetic energy of the macroscopic motion) plus Maxwellian magnetic energy is equal to the gravitational energy plus the Proca part of the magnetic energy. The bound in Eq. \ref{bound} assumes that the virial theorem can be applied, approximately, to the random part of the galactic motion, after the mean rotation of the galaxy has been excluded.


{\bf CONCLUSIONS: } When trying to measure $m$ one must distinguish between measurements performed on large and small scales. If the photon acquires mass by the Higgs mechanism, the large-scale behavior of the photon might be effectively Maxwellian. If, on the other hand, one postulates the Proca regime for all scales, the very existence of the galactic field implies $m<10^{-26}$ eV, as correctly calculated by Yamaguchi and Chibisov \cite{yam, chibisov}.

We thank S. Dimopoulos, G. Farrar, M. Shaposhnikov, A. Vilenkin for useful discussions. 
E.A. thanks the NYU Cosmology and Particle Physics Group for a stimulating visit. G.D. and A.G. are supported by the David and Lucile  Packard Foundation,
and G.D. is also supported by the Alfred P. Sloan Foundation and by NSF PHY-0070787.

\end{document}